\PassOptionsToPackage{table}{xcolor} 
\documentclass{article} 
\usepackage{iclr2027_conference,times}

\usepackage{amsmath,amsfonts,bm}

\def\eqref#1{equation~\ref{#1}}

\def\1{\bm{1}}

\DeclareMathAlphabet{\mathsfit}{\encodingdefault}{\sfdefault}{m}{sl}
\SetMathAlphabet{\mathsfit}{bold}{\encodingdefault}{\sfdefault}{bx}{n}

\usepackage[table]{xcolor}
\definecolor{AweAIblue}{HTML}{386B58}
\usepackage[colorlinks=true, linkcolor=AweAIblue, citecolor=AweAIblue, urlcolor=AweAIblue, hypertexnames=false]{hyperref}
\usepackage{url}
\usepackage[utf8]{inputenc} 
\usepackage[T1]{fontenc}    
\usepackage{url}            
\usepackage{booktabs}       
\usepackage{threeparttable}
\usepackage{tabularx}
\usepackage{ragged2e}
\usepackage{amsfonts}       
\usepackage{bbm}            
\usepackage{nicefrac}       
\usepackage{microtype}      
\usepackage{xcolor}         
\usepackage{graphicx}
\usepackage{mathrsfs}
\usepackage{amsmath}
\usepackage{subfigure}      

\usepackage{footnote}
\usepackage{footnotebackref}
\usepackage{multirow}
\usepackage{makecell}
\usepackage{amsmath}

\usepackage{float}  
\usepackage{subfigure}  
\usepackage{tikz}
\usetikzlibrary{shapes,arrows}
\usepackage{appendix}
\usepackage{bm}
\usepackage{amsbsy}
\usepackage{pifont}
\usepackage{graphicx} 
\usepackage{wrapfig}
\usepackage{bbding}
\usepackage{titlesec}
\usepackage{colortbl}
\usepackage{subcaption}
\usepackage[table]{xcolor}
\usepackage{hyperref} 
\usepackage{etoc}     
\usepackage{enumitem}
\usepackage{xparse}

\usepackage[most]{tcolorbox}
\usepackage{listings}

\definecolor{S2Sage}{HTML}{4F8071}
\definecolor{S2Paper}{HTML}{F7F3EA}
\definecolor{S2Line}{HTML}{D6DED7}

\newtcblisting{PromptBox}[2][]{
    enhanced jigsaw,
    breakable,
    listing only,
    listing options={
        basicstyle=\footnotesize\ttfamily,
        breaklines=true,
        breakatwhitespace=false,
        columns=fullflexible,
        keepspaces=true,
        aboveskip=0pt, belowskip=0pt,
        breakindent=0pt,
        breakautoindent=false,
        upquote=true,
        literate=
            {"}{\char`\"}1
            {'}{\char`\'}1
            {"}{\char`\"}1
            {'}{\char`\'}1
            {…}{...}1
            {—}{---}1
            {–}{--}1,
    },
    colframe=S2Line,
    colback=S2Paper,
    coltitle=white,
    colbacktitle=S2Sage,
    title={#2},
    fonttitle=\large\bfseries,
    arc=4mm,
    boxsep=3pt,
    top=2pt, bottom=2pt,
    #1
}

\definecolor{purple}{HTML}{c994c7}
\definecolor{navyblue}{RGB}{30,130,255}
\definecolor{citecolor}{HTML}{386B58}
\definecolor{lightgray}{gray}{0.9}
\definecolor{blanchedalmond}{rgb}{1.0, 0.92, 0.8}
\definecolor{cerise}{rgb}{0.871, 0.192, 0.388}
\definecolor{DarkNature}{HTML}{1B7837}
\definecolor{LightNature}{HTML}{7FBF7B}
\definecolor{BestResultBG}{HTML}{DFF0E3}
\definecolor{SecondResultBG}{HTML}{F0F7EF}

\definecolor{TaskBG}{HTML}{EFE6FF}        
\definecolor{StateBG}{HTML}{F5F5F7}       
\definecolor{ExpertBG}{HTML}{EAF7EA}      
\definecolor{IWMBG}{HTML}{FDECF3}         
\definecolor{SRBG}{HTML}{E6F2FF}          

\newcolumntype{L}[1]{>{\RaggedRight\arraybackslash}p{#1}} 
\newcolumntype{Y}{>{\RaggedRight\arraybackslash}X}        

\newcommand{\header}[1]{\vspace*{1mm}\noindent{\textbf{#1}}}
\newcommand{\model}{E\textsuperscript{2}Sim}
\newcommand{\actscore}{ACT}
\newcommand{\dinglabelone}{\textcolor[HTML]{D98282}{\ding{182}}}
\newcommand{\dinglabeltwo}{\textcolor[HTML]{6E9B72}{\ding{183}}}
\newcommand{\dinglabelthree}{\textcolor[HTML]{6B8FD6}{\ding{184}}}

\definecolor{AblationBlue}{HTML}{E1EBFE}
\definecolor{AblationGreen}{HTML}{DCEADA}
\definecolor{AblationPink}{HTML}{F9D5D4}
\newcommand{\ablationdot}[1]{%
  \tikz[baseline=-0.55ex]\filldraw[fill=#1,draw=black!45,line width=0.35pt]
  (0,0) circle (0.72ex);%
}

\title{Adaptive Resource Allocation for Effective and Efficient LLM Social Survey Simulation}

\author{Yuanzi Li, Xueyang Feng, Junhao Wang, Lei Wang, Xu Chen\thanks{Corresponding author.}\\
Renmin University of China\\
{\fontfamily{pcr}\selectfont liyuanzi0313@outlook.com, xu.chen@ruc.edu.cn}
}

\iclrfinalcopy 
\begin{document}

\maketitle

\ificlrfinal
    \lhead{Published as a conference paper at ICLR 2027}
\else
    \lhead{Under review as a conference paper at ICLR 2027}
\fi

\begin{abstract}
Large Language Models (LLMs) enable scalable social survey simulation, yet existing pipelines typically assign the same strong general-purpose model and a fixed, often large, respondent history to every respondent--question request. This uniform design overlooks three properties of survey simulation. 
\textbf{\emph{(1) Model capability.}}
A stronger general-purpose model may rely more on its own knowledge and deviate from respondent-specific evidence, while also costing more.
\textbf{\emph{(2) Respondent history.}}
More history can help when evidence is insufficient, but irrelevant responses may introduce noise and lengthen the input.
\textbf{\emph{(3) Model--history interaction.}}
Changing the history budget can change the preferred model, while changing the model can change the preferred history budget.
These considerations call for joint request-level allocation under an explicit accuracy--cost trade-off. We propose \textbf{\emph{\model}}, an adaptive resource allocation framework that treats each model--history configuration as a joint request-level decision. Conditioned on the respondent persona, ranked response history, and target question, a lightweight policy predicts the accuracy--cost score of every configuration and selects the most suitable one. To reduce sensitivity to incomplete or variable histories, we apply respondent-history drop-and-swap augmentation; to handle differences in allocation ambiguity, we use a margin-based curriculum that progresses from clearly separated configurations to harder cases. Experiments on four real-world social-survey datasets, multiple model pools, and different history-budget spaces demonstrate improvements over oracle best-fixed configurations, with gains of up to 6.3 percentage points in Accuracy and reductions of up to 55.3\% in cost. 
Our code is available at \url{https://anonymous.4open.science/r/E2Sim-DE66}.
\end{abstract}

\section{Introduction}

Social surveys are the empirical backbone of computational social science: they record how attitudes, values, and policy preferences are distributed across a heterogeneous population \citep{squazzoni2020computational,hamill2009social,wang2025user}. Such evidence, however, is hard to scale and expensive to obtain, since fielding a questionnaire repeatedly and at scale requires substantial human effort \citep{wright2010survey,heffetz2019difficulty,kalton2009methods,groves2011survey,roopa2012questionnaire,tourangeau2000psychology}. Large Language Models (LLMs)~\citep{minaee2024large,achiam2023gpt} offer a scalable complement: conditioned on a respondent's demographic profile and past answers, an LLM can simulate how that specific person would answer a new question \citep{argyle2023out,aher2023using,cao2023assessing,durmus2023towards}. Given a target question the respondent has not answered, a simulator retrieves the $k$ past responses of that respondent most relevant to the question, which serve as personal references that ground the prediction in the respondent's own expressed views, and prompts a model $m$ with the respondent's persona, these $k$ responses, and the question to predict the answer. For example, for the question \emph{``Is the government spending too much, too little, or about the right amount on healthcare?''}, the input is a retired respondent's persona together with that respondent's retrieved earlier answers on government responsibility and on medical expenses, and the output is one of the given options.

\begin{figure*}[!t]
    \centering
    \includegraphics[width=\textwidth]{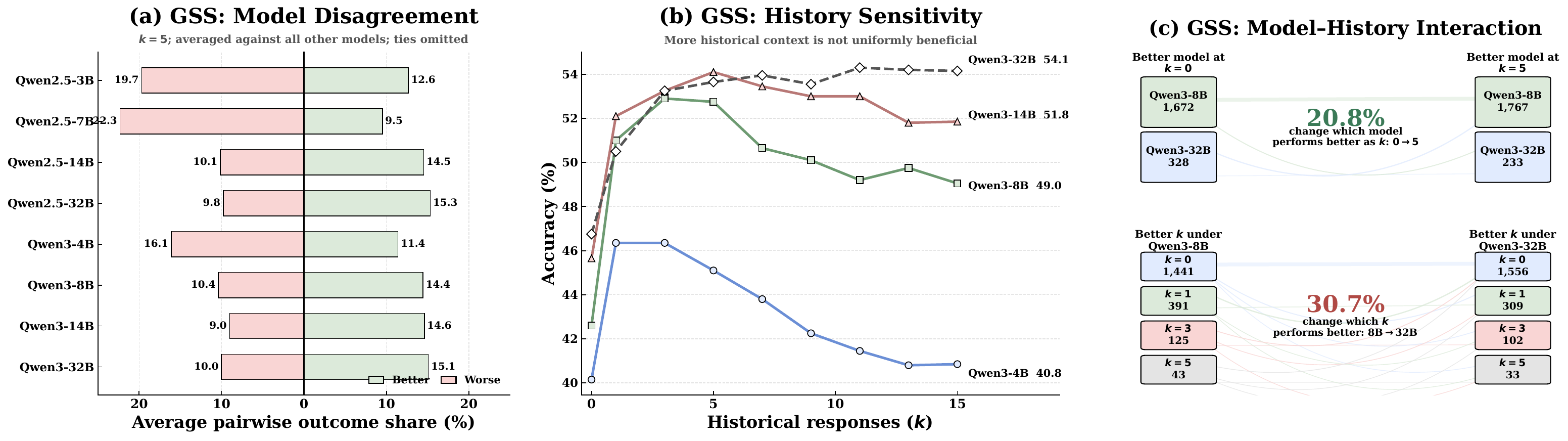}
    \vspace{-1.5em}
    \caption{Analysis of 2,000 GSS respondent--question requests. (a) Proportions of pairwise wins and losses for each model. (b) Accuracy as the history budget varies. (c) Model--history interaction: 20.8\% of requests change their better-performing model as $k$ increases from 0 to 5, while 30.7\% change their better-performing $k$ when switching from Qwen3-8B to Qwen3-32B.}
    \label{fig:preexp}
    \vspace{-2em}
\end{figure*}

However, existing LLM-based survey simulation pipelines typically use the same high-capability general-purpose model and a fixed, often large, respondent history for every respondent--question request. This one-size-fits-all assumption overlooks two forms of heterogeneity and their interaction:
\textbf{\emph{(1) Model capability (m).}}
Different survey requests require different levels of model capability.
A stronger general-purpose model is not always better for survey simulation, because the goal is to predict a respondent's likely choice from limited personal evidence rather than rely solely on general knowledge to derive the most reasonable answer.
Requests with sparse or conflicting evidence may benefit from stronger reasoning capabilities. However, when the respondent's pattern is clear, a smaller model may already suffice, while a stronger model may introduce additional priors that are less aligned with the respondent-specific pattern and incur higher inference cost.
As shown in Figure~\ref{fig:preexp}(a), no single model performs best across all requests, motivating adaptive selection of $m$ at the request level rather than using one fixed model for every request.
\textbf{\emph{(2) Respondent history (k).}}
Different survey requests require different amounts of respondent history.
We rank a respondent's historical responses by their relevance to the target question and include the most relevant ones first.
If the most relevant responses are insufficient, increasing $k$ may provide useful evidence. However, once the available history is sufficient, further increases mainly introduce lower-ranked responses that may add little value or distract the model, while increasing prompt length and inference cost.
As shown in Figure~\ref{fig:preexp}(b), larger $k$ can yield diminishing or even negative returns, motivating adaptive selection of the history budget rather than using one fixed value for every request.
\textbf{\emph{(3) Model--history interaction (m, k).}}
Model choice and history budget are interdependent.
For the same request, changing the model $m$ may change the most suitable history budget $k$, because models differ in how effectively they extract useful signals from respondent history.
Conversely, changing the history budget $k$ may change the most suitable model $m$, because the amount and quality of available evidence affect how much model capability is useful.
As shown in Figure~\ref{fig:preexp}(c), increasing $k$ from 0 to 5 changes the better-performing model for 20.8\% of requests. Switching from Qwen3-8B to Qwen3-32B changes the better-performing history budget for 30.7\% of requests.
This interaction motivates selecting $(m,k)$ jointly at the request level rather than choosing the model and history budget independently.


Motivated by these observations, we propose \textbf{\model}, an adaptive resource allocation framework for \underline{\textbf{E}}ffective and \underline{\textbf{E}}fficient LLM social survey \underline{\textbf{Sim}}ulation. \model\ treats each respondent--question pair as a distinct allocation unit. We score each candidate configuration by its Accuracy--Cost Trade-off (\actscore), $\actscore=\mathrm{Accuracy}-\alpha\cdot\mathrm{Cost}$, in which the hyperparameter $\alpha$ sets how heavily inference cost is penalized relative to a correct prediction. Using offline evaluations of candidate configurations as supervision, a lightweight policy learns to predict their expected \actscore\ from the respondent persona, ranked response history, and target question, and jointly selects $(m,k)$ at inference time. Learning this policy raises two challenges. The first is that the retrieved history is not fixed: the same respondent may have fewer responses available in another setting, and a small change in retrieval similarity can change their order. If the policy is trained on a single version of the history, it may tie its decision to that exact version and choose a different configuration when one response is missing or two are reordered. We therefore train on perturbed histories, dropping and swapping retrieved responses, so that the policy gives the same request a consistent allocation and decides from how informative the history is as a whole. The second is that requests are not equally easy to allocate. When the history clearly points to an answer, one configuration stands out; when the evidence is sparse or self-contradictory, several configurations perform almost identically and the supervision is close to a tie. We quantify this difficulty by the \actscore\ margin between the best two configurations, and train with a curriculum that starts from large-margin requests and gradually adds ambiguous ones.

\noindent\textbf{Contributions.}
\dinglabelone\ We identify complementary, request-dependent roles of model capability and respondent history, and show that their benefits are interdependent.
\dinglabeltwo\ We propose \model, a cost-aware joint allocation framework with respondent-history augmentation and a margin-based curriculum for robust policy learning.
\dinglabelthree\ Across four surveys, we evaluate \model\ under varied model pools and history budgets, demonstrating its effectiveness and robustness.

\section{Problem Formulation and Definition}
\label{sec:problem_formulation}


\textbf{Formulation: LLM-based Social Survey Simulation.}
Let $u$ denote a respondent with a textual persona $P_u$, and let $S(u)$
denote the respondent's observed responses to other survey questions. For a
target question $q$, including its response options, a simulation request is
defined as
\begin{equation}
    x=(P_u,S(u),q).
\end{equation}
The amount of respondent history provided as personalized context is
controlled by a history budget $k$. Given $k$, the simulator retrieves an
ordered context $H_k$ containing the $k$ historical responses most relevant
to the target question, and model $m$ predicts the respondent's answer:
\begin{equation}
    H_k=\operatorname{Retrieve}_k\!\left(q,S(u)\right),\qquad
    \hat{y}_{m,k}\sim
    P_m\!\left(\cdot\mid P_u\oplus H_k\oplus q\right),
\end{equation}
where $\oplus$ denotes prompt concatenation.

\textbf{Definition: Adaptive Resource Allocation.}
Let $\mathcal{M}$ be the candidate model pool and $\mathcal{K}$ the available
history budgets. Their Cartesian product defines the joint action space
\begin{equation}
    \mathcal{A}=\mathcal{M}\times\mathcal{K},
    \qquad a=(m,k)\in\mathcal{A}.
\end{equation}
An allocation policy maps each request to one joint action,
\begin{equation}
    \pi:\mathcal{X}\rightarrow\mathcal{A},
    \qquad \pi(x)=(m,k).
\end{equation}
This joint formulation captures the interdependence between model choice and
history budget: the relative performance of candidate models can change with
$k$ for a given respondent--question request, while the predictive value of
additional history can likewise change with $m$.

\section{Method}
\label{sec:method}

\begin{figure*}[!t]
    \centering
    \includegraphics[width=\textwidth]{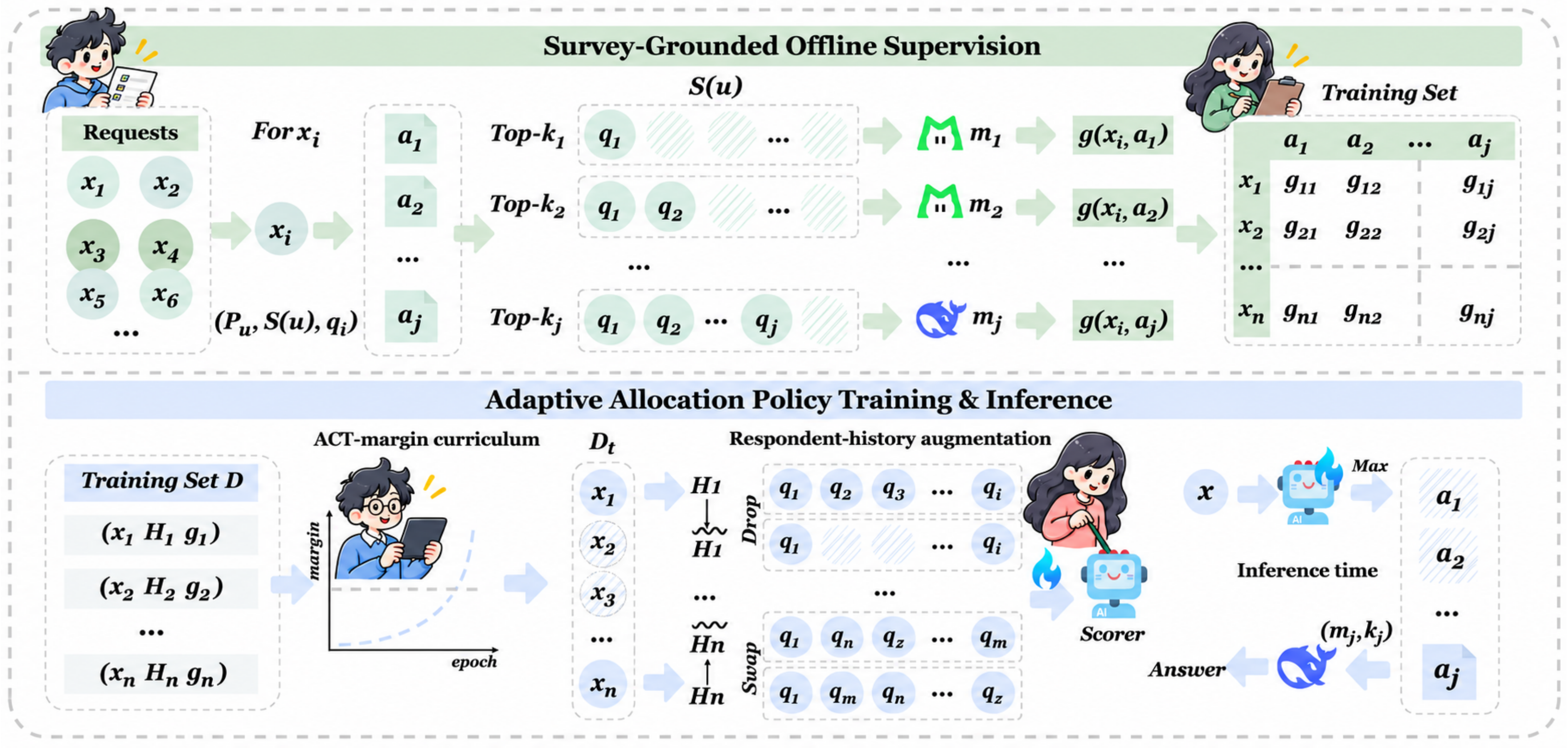}
    \vspace{-1em}
    \caption{Overview of \model. The upper panel shows the construction of survey-grounded supervision: it retrieves target-relevant respondent history, evaluates every model--history action, and records the complete \actscore\ vector. The lower panel trains a joint configuration scorer with respondent-history Drop/Swap and an \actscore-margin curriculum; at inference, the policy invokes only its selected $(m,k)$.}
    \vspace{-1em}

    \label{fig:E2Sim-framework}
\end{figure*}

In this section, we present \model, the adaptive resource allocation framework illustrated in Figure~\ref{fig:E2Sim-framework}. We first construct survey-grounded offline supervision by evaluating every candidate model--history configuration against observed survey responses (Section~\ref{sec:offline_data}). We then learn a joint model--history allocation policy with respondent-history augmentation and an \actscore-margin curriculum (Section~\ref{sec:allocation_policy}). At deployment, request-level adaptive inference selects one model and history budget for each request and invokes the chosen configuration (Section~\ref{sec:inference}).

\subsection{Survey-Grounded Offline Supervision}
\label{sec:offline_data}

Each training request $x_i$ is paired with the response $y_i$ recorded in the
survey. We run every action $a=(m,k)$ on the same request, using model $m$ and
the top-$k$ historical responses relevant to its target question. The generated
answer, input-token usage and model price determine the Accuracy, Cost, and \actscore\ of
that action. Collecting these scores across $\mathcal{A}$ gives one complete
supervision vector $\mathbf{g}_i$ for learning which configuration suits the
request.

Respondent histories span many social topics, so the context is formed relative
to the target question. Let $S(u_i)$ be respondent $u_i$'s observed response
history. A fixed embedding model ranks its questions by cosine similarity and
returns the top-$k$ question--answer pairs:
\begin{equation}
    \mathbf{e}(q)=\operatorname{Embed}(q),\quad
    s(q_i,q)=\frac{\mathbf{e}(q_i)^{\top}\mathbf{e}(q)}
    {\lVert\mathbf{e}(q_i)\rVert_2\,\lVert\mathbf{e}(q)\rVert_2},\quad
    H_{i,k}=\operatorname{TopK}_{\substack{(q',y')\in S(u_i)\\q'\neq q_i}}
    s(q_i,q').
\label{eq:topk_retrieval}
\end{equation}
For each action $a=(m,k)$, model $m$ receives the persona, retrieved
history, and target question. Its output and exact-match accuracy against the
observed response $y_i$ are
\begin{equation}
    \hat{y}_i(a)\sim
    P_m\!\left(\cdot\mid P_{u_i}\oplus H_{i,k}\oplus q_i\right),\qquad
    \operatorname{Acc}(x_i,a)=
    \mathbbm{1}\!\left[\hat{y}_i(a)=y_i\right].
\label{eq:action_response}
\end{equation}

Because different actions trade predictive accuracy against inference cost, we define a unified score that captures both.  
The cost of an action depends on two factors: model choice determines the token price, while the history budget determines the prompt length.  
Since the structured output contains only an option index, we measure inference cost using only price-weighted input tokens.  
To make cost comparable with accuracy, we normalize it to the $[0,1]$ scale and define
\begin{equation}
C(x_i,a)=
\frac{c_m^{\mathrm{in}}T(x_i,a)}
{\displaystyle\max_{1\leq j\leq N,\;a'=(m',k')\in\mathcal{A}}
c_{m'}^{\mathrm{in}}T(x_j,a')},\qquad
g_i(a)=\operatorname{Acc}(x_i,a)-\alpha C(x_i,a),
\label{eq:normalized_cost}
\end{equation}
where $T(x_i,a)$ is the number of input tokens, $c_m^{\mathrm{in}}$ is the model-specific input-token price, and $g_i(a)$ is the Accuracy--Cost Trade-off score (\actscore).  
The coefficient $\alpha$ controls the trade-off between accuracy and cost.  
Evaluating every action gives the complete score vector
$\mathbf{g}_i=(g_i(a))_{a\in\mathcal{A}}$.  
With $H_i=H_{i,k_{\max}}$ and $k_{\max}=\max\mathcal{K}$, the offline supervision is
\begin{equation}
    \mathcal{D}
    =\left\{
    \left(x_i,H_i,\mathbf{g}_i\right)
    \right\}_{i=1}^{N}.
    \label{eq:offline_training_set}
\end{equation}
Thus, each request retains the relative accuracy--cost outcomes of every
model--history alternative rather than only a best-action label.

\subsection{Joint Model--History Allocation Policy}
\label{sec:allocation_policy}

\textbf{Joint configuration scoring.}
Each survey request provides three complementary signals for allocation: the
persona describes relatively stable sociodemographic characteristics, the
respondent history records issue-specific choices, and the target question
specifies the attitude to be inferred. We combine them into the request
representation
\begin{equation}
    z_i=P_{u_i}\oplus H_i\oplus q_i,
    \label{eq:policy_input}
\end{equation}
where $\oplus$ denotes concatenation and $H_i=H_{i,k_{\max}}$ is the context
under the predefined maximum history budget
$k_{\max}=\max\mathcal{K}$. Providing the policy with the maximum available history gives it sufficient information to compare configurations with different values of $k$. A lightweight scorer $f_\theta$ then predicts
the \actscore\ of every joint action in one pass:
\begin{equation}
    f_\theta(z_i)
    =\widehat{\mathbf{g}}_\theta(z_i)
    =\bigl(\widehat{g}_\theta(z_i,a)\bigr)_{a\in\mathcal{A}}
    \in\mathbb{R}^{|\mathcal{A}|}.
    \label{eq:policy_output}
\end{equation}
Each coordinate corresponds to one model--history configuration and is
supervised by its matching entry in $\mathbf{g}_i$. Joint scoring lets the same
survey evidence determine the predicted values of both $m$ and $k$, preserving
their interaction within a single allocation decision.

\textbf{Robustness to incomplete and variable histories.}
In practice, response histories may be incomplete due to item nonresponse, while small differences in retrieval similarity can alter the ordering of otherwise comparable responses.
If the policy relies on memorizing whether a response appears or where it appears in the history, its allocations may become unstable.
To improve robustness, we construct alternative views of the same request by dropping and swapping responses:
\begin{equation}
    \widetilde{H}_i
    =\operatorname{Swap}_{p_{\mathrm{swap}}}
    \!\left(\operatorname{Drop}_{p_{\mathrm{drop}}}(H_i)\right),\qquad
    \widetilde{z}_i=P_{u_i}\oplus\widetilde{H}_i\oplus q_i.
    \label{eq:fewshot_augmentation}
\end{equation}
where $p_{\mathrm{drop}}$ is the probability of independently removing each
response, and $p_{\mathrm{swap}}$ is the probability of exchanging the
positions of two selected responses. These limited perturbations preserve the
respondent persona and target question while varying only the retrieved
history, yielding alternative views of the same request. Training on these
views acts as a local invariance regularizer, encouraging stable allocation
under small changes in history content and order.

\textbf{\actscore-margin curriculum for allocation ambiguity.}
Survey requests differ in how clearly their available evidence identifies an
appropriate resource configuration. Target-relevant history may sharply
separate one action from the alternatives, whereas sparse or mixed evidence
can leave several actions nearly tied. The fully observed score vector makes
this allocation ambiguity directly measurable. Let $g_i^{(1)}$ and
$g_i^{(2)}$ be the largest and second-largest action scores; their margin is
\begin{equation}
    \Delta_i=g_i^{(1)}-g_i^{(2)}.
    \label{eq:score_margin}
\end{equation}
A large $\Delta_i$ indicates a clear allocation preference, while a small
margin indicates ambiguity. Rather than exposing the policy to both from the
outset, we begin with large-margin requests and expand the admitted fraction.
At epoch $t$, the fraction $\rho_t$ and corresponding training subset
$\mathcal{D}_t$ are
\begin{equation}
    \rho_t=
    \begin{cases}
    \displaystyle
    \rho_0+(1-\rho_0)\frac{t-1}{T-1}, & 1\leq t\leq T,\\[2mm]
    1, & t>T,
    \end{cases}
    \qquad
    \mathcal{D}_t=
    \operatorname{Top}_{\lceil \rho_t N\rceil}
    \left(\{x_i\}_{i=1}^{N};\,\Delta_i\right).
    \label{eq:curriculum_schedule}
\end{equation}
where $\rho_0$ is the initial fraction, $T$ is the number of curriculum
epochs, and $\mathcal{D}_t$ contains the requests with the largest margins.
Thus, the policy first learns from requests with clear resource requirements
and progressively incorporates more ambiguous allocation decisions as
$\rho_t$ increases.

\textbf{Joint score regression.}
A best-action label would discard how strongly one configuration is preferred:
a narrow and a decisive \actscore\ gap would receive identical supervision.
This distinction matters for survey requests whose allocation ambiguity varies.
We therefore regress the complete score vector. At curriculum epoch $t$, the
Smooth-$L_1$ objective is
\begin{equation}
    \mathcal{L}(\theta)
    =\frac{1}{|\mathcal{D}_t|}
    \sum_{x_i\in\mathcal{D}_t}
    \ell_{\mathrm{smooth}\text{-}L_1}\!\left(
    \widehat{\mathbf{g}}_\theta(\widetilde{z}_i),
    \mathbf{g}_i\right).
    \label{eq:score_regression_loss}
\end{equation}
Each coordinate is supervised by the observed score of the corresponding
model--history action, preserving both the ranking and the utility differences
among all candidate configurations.

\subsection{Request-Level Adaptive Inference}
\label{sec:inference}

At deployment, each respondent--question request is allocated independently.
We retrieve the predefined maximum history once and form
$z(x)=P_u\oplus H_{k_{\max}}(u,q)\oplus q$. The policy directly selects
\begin{equation}
    (m^*,k^*)=
    \arg\max_{(m,k)\in\mathcal{A}}
    \widehat{g}_\theta\!\left(z(x),(m,k)\right).
    \label{eq:inference_policy_input}
\end{equation}
The policy examines up to \(k_{\max}\) responses and selects a survey model \(m^*\) and history budget \(k^*\) for each request. The selected model \(m^*\) then predicts the answer using only the top \(k^*\) responses, so the additional history examined by the policy does not increase the survey model's prompt length. At deployment, each request requires one policy evaluation and one call to the selected survey model.

\section{Experiments}
\label{sec:experiments}

We investigate four research questions. \hyperref[sec:rq1_main]{\textbf{(RQ1)}} How effectively does \model\ improve the accuracy--cost trade-off, and how quickly can its offline supervision cost be amortized? \hyperref[sec:rq2_tradeoff]{\textbf{(RQ2)}} How does $\alpha$ control the cost--accuracy trade-off? \hyperref[sec:rq3_robustness]{\textbf{(RQ3)}} How robust is \model\ across different history-budget spaces and model-pool sizes? \hyperref[sec:rq4_ablation]{\textbf{(RQ4)}} How much does each component contribute?

\subsection{Experimental Setup}

\header{Datasets.} Experiments are conducted on four real-world social-survey datasets: \textbf{ANES} (American National Election Studies, 2024 wave), \textbf{BSA} (British Social Attitudes, 2024 wave), \textbf{GSS} (General Social Survey, 2024 wave), and \textbf{WVS} (World Values Survey, Wave 7). Dataset statistics and split construction are detailed in Appendix~\ref{app:data_splits}.

\header{Model and History Configuration.} The candidate model pool consists of three LLMs: Qwen3-8B, Qwen3-32B, and DeepSeek-V4-Flash. These models span a range of inference costs and reasoning capabilities. For our controlled cost model, we assign relative input-token weights of $2$, $4$, and $4$ to Qwen3-8B, Qwen3-32B, and DeepSeek-V4-Flash, respectively. These weights are used consistently in the normalized Cost and \actscore\ calculations. The history-budget space is $\mathcal{K} = \{0, 1, 3, 5\}$, where $k$ specifies the number of historical responses and $k=0$ denotes the no-history setting. 


\header{Baselines.} We compare \model\ with four baselines: (1) Worst-Case Reference (WCR), which has access to the observed action scores and selects the configuration with the lowest \actscore\ for each request, providing a worst-case reference; (2) Random, which samples a model--history configuration uniformly from $\mathcal{A}$; (3) LLM-based Selector (LLM-S), which uses Qwen3-32B with the prompt in Appendix~\ref{sec:llm_selector_prompt} to predict a suitable configuration, with the associated Qwen3-32B inference cost included in the reported Cost; and (4) Oracle Best Fixed (BF), which uses the observed action scores on the evaluation split to select the single configuration with the highest average \actscore\ and applies it to every request. BF is therefore an oracle upper bound for policies restricted to one fixed configuration.

\header{Evaluation Metrics.} We report three metrics: (1) Accuracy, the proportion of requests where the predicted answer matches the ground truth; (2) Cost, the average price-weighted input-token cost normalized by the maximum weighted cost among all candidate configurations for all requests; and (3) Accuracy--Cost Trade-off score (\actscore), the combined metric $\text{Acc} - \alpha \cdot \text{Cost}$ that balances effectiveness and efficiency, where higher Accuracy and \actscore\ and lower Cost indicate better performance.

\definecolor{positive}{HTML}{3F8F5F}   
\definecolor{negative}{HTML}{757575}   
\definecolor{neutral}{HTML}{757575}    
\providecommand{\posresult}[1]{\textcolor{positive}{\textbf{#1}}}
\providecommand{\negresult}[1]{\textcolor{negative}{\textbf{#1}}}
\providecommand{\neuresult}[1]{\textcolor{neutral}{\textbf{#1}}}

\begin{table*}[!t]
  \centering
  \caption{Overall performance under two model configurations. Improved reports the difference between \model\ and the strongest baseline by ACT.}
  \label{tab:main}
  \setlength{\tabcolsep}{4.2pt}
  \renewcommand{\arraystretch}{1.08}
  \resizebox{\textwidth}{!}{%
    \begin{tabular}{ll!{\vrule width 0.6pt}cccccc!{\vrule width 0.6pt}cccccc}
      \toprule
      \multirow{2}{*}{\textbf{Dataset}}
      & \multirow{2}{*}{\textbf{Metric}}
      & \multicolumn{6}{c!{\vrule width 0.6pt}}{\textbf{Qwen3-8B + Qwen3-32B}}
      & \multicolumn{6}{c}{\textbf{Qwen3-8B + DS-V4-Flash}} \\
      \cmidrule(lr){3-8}\cmidrule(lr){9-14}
      & & \textbf{WCR} & \textbf{Random} & \textbf{LLM-S} & \textbf{BF} & \textbf{Ours} & \textbf{Improved}
      & \textbf{WCR} & \textbf{Random} & \textbf{LLM-S} & \textbf{BF} & \textbf{Ours} & \textbf{Improved} \\
      \midrule
      \multirow{3}{*}{ANES}
      & Accuracy & 0.2047 & 0.5737 & 0.6360 & 0.6552 & 0.6880 & \posresult{0.0328} & 0.2097 & 0.5880 & 0.6585 & 0.6850 & 0.7063 & \posresult{0.0213} \\
      & Cost & 0.5559 & 0.3917 & 1.5882 & 0.5736 & 0.3614 & \posresult{0.2122} & 0.5399 & 0.3910 & 1.6052 & 0.5690 & 0.3577 & \posresult{0.2113} \\
      & ACT & 0.1770 & 0.5542 & 0.5566 & 0.6266 & 0.6699 & \posresult{0.0433} & 0.1828 & 0.5684 & 0.5782 & 0.6565 & 0.6884 & \posresult{0.0319} \\
      \midrule
      \multirow{3}{*}{BSA}
      & Accuracy & 0.1092 & 0.4275 & 0.4675 & 0.4825 & 0.5102 & \posresult{0.0277} & 0.1095 & 0.4472 & 0.4843 & 0.5155 & 0.5222 & \posresult{0.0067} \\
      & Cost & 0.5871 & 0.3848 & 1.5262 & 0.5680 & 0.3870 & \posresult{0.1810} & 0.5692 & 0.3831 & 1.5458 & 0.5620 & 0.3706 & \posresult{0.1914} \\
      & ACT & 0.0799 & 0.4083 & 0.3912 & 0.4541 & 0.4909 & \posresult{0.0368} & 0.0810 & 0.4281 & 0.4070 & 0.4874 & 0.5037 & \posresult{0.0163} \\
      \midrule
      \multirow{3}{*}{GSS}
      & Accuracy & 0.1787 & 0.5085 & 0.5242 & 0.5278 & 0.5543 & \posresult{0.0265} & 0.1872 & 0.5210 & 0.5385 & 0.5785 & 0.5790 & \posresult{0.0005} \\
      & Cost & 0.5445 & 0.3582 & 1.3959 & 0.3205 & 0.2873 & \posresult{0.0332} & 0.5272 & 0.3556 & 1.3874 & 0.6370 & 0.4738 & \posresult{0.1632} \\
      & ACT & 0.1515 & 0.4906 & 0.4545 & 0.5117 & 0.5399 & \posresult{0.0282} & 0.1609 & 0.5032 & 0.4691 & 0.5466 & 0.5553 & \posresult{0.0087} \\
      \midrule
      \multirow{3}{*}{WVS}
      & Accuracy & 0.1888 & 0.4973 & 0.5407 & 0.5533 & 0.5687 & \posresult{0.0154} & 0.1935 & 0.5258 & 0.5447 & 0.5935 & 0.5998 & \posresult{0.0063} \\
      & Cost & 0.4935 & 0.3138 & 1.2424 & 0.6050 & 0.2704 & \posresult{0.3346} & 0.4853 & 0.3188 & 1.2817 & 0.4710 & 0.3457 & \posresult{0.1253} \\
      & ACT & 0.1641 & 0.4816 & 0.4786 & 0.5230 & 0.5552 & \posresult{0.0322} & 0.1692 & 0.5098 & 0.4807 & 0.5700 & 0.5825 & \posresult{0.0125} \\
      \bottomrule
    \end{tabular}%
  }
  \vspace{-0.6em}
\end{table*}

\header{Implementation Details.} Question embeddings use OpenAI's
\texttt{text-embedding-3-small}; LLM responses use temperature 0 with one evaluation per request--configuration.
The allocation policy is implemented with a
ModernBERT-base encoder followed by a linear scoring head. Each input is truncated to 1,250 tokens. The model is optimized with AdamW using a learning rate of
$1\times10^{-4}$, a 0.1 linear warmup ratio, and a batch
size of 12. The complete training schedule uses 10
curriculum epochs followed by 10 additional full-data epochs. During the
curriculum phase, the initial training fraction is 0.3 and increases linearly
to include all requests. For respondent-history augmentation, Drop is applied with
probability 0.25, removing at most two responses and at most 40\% of the
context, while Swap is applied with probability 0.15. The lightweight
ModernBERT-base policy model (149M parameters) is deployed locally; we price it at $2\times149\text{M}/8\text{B}\approx0.0373$ on the same relative scale and include this cost in the reported Cost for \model, alongside the LLM-based Selector's Qwen3-32B overhead (Appendix~\ref{sec:cost_computation}). All reported \model\ results are averaged over three training seeds (41, 42, and 43).
Further implementation details are available in our released code at
\url{https://anonymous.4open.science/r/E2Sim-DE66}.

\subsection{Main Results}
\label{sec:rq1_main}

\textbf{Overall Performance.} Table~\ref{tab:main} presents the performance of \model\ and four baselines across four social-survey datasets and two model configurations, using $\mathcal{K}=\{0,1,3,5\}$ and $\alpha=0.05$. \model\ outperforms both Random and LLM-based Selector (LLM-S) in terms of \actscore\ across all datasets. More importantly, \model\ also improves upon Oracle Best Fixed (BF), the evaluation-set oracle upper bound over fixed configurations. Under Qwen3-8B + Qwen3-32B, \model\ improves Accuracy by 1.5--3.3 percentage points, reduces normalized Cost by 10.4--55.3\%, and raises \actscore\ by 2.8--4.3 percentage points across the four datasets. Under Qwen3-8B + DeepSeek-V4-Flash, \model\ improves Accuracy by 0.1--2.1 percentage points, reduces Cost by 25.6--37.1\%, and raises \actscore\ by 0.9--3.2 percentage points relative to BF on all four datasets. These results show that adaptive model--history allocation consistently improves the effectiveness--efficiency objective across all four datasets and both model configurations, with especially substantial savings in inference cost.

\begin{figure*}[t]
    \centering
    \includegraphics[width=\textwidth]{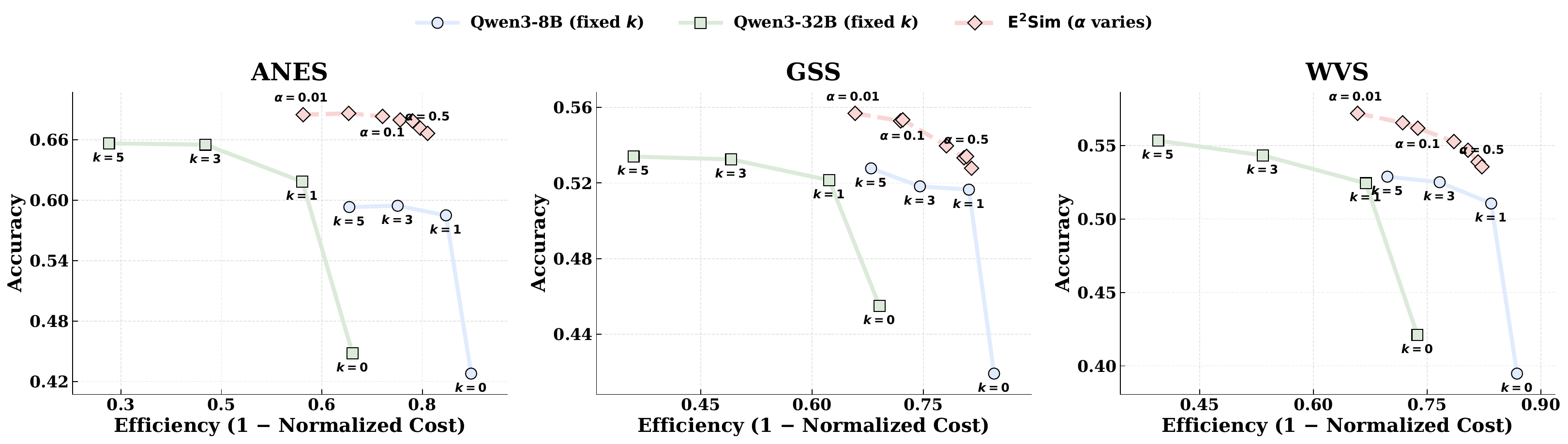}
    \vspace{-1em}
    \caption{Cost-accuracy tradeoff under different alpha values. The plot shows fixed baselines (Qwen3-8B and Qwen3-32B with different history budgets $k$) and ours at various alpha values, demonstrating the trade-off between cost efficiency and accuracy.}
    \vspace{-1em}
    \label{fig:cost_accuracy_tradeoff}
\end{figure*}

\header{Offline-Supervision Amortization.} We calculate the cost of collecting offline supervision by summing the price-weighted token costs of evaluating eight actions for each of 16,000 training requests. Because \model\ improves Accuracy while reducing Cost, we amortize this one-time expense using its per-request \actscore\ gain over BF. Table~\ref{tab:offline_amortization} shows break-even points of 57K--78K requests across the four datasets, only 3.6--4.9 times the training volume. WVS reaches break-even after 62K requests, or 485 respondent-equivalents at the observed density of 127.4 requests per respondent; its original population of 97,220 respondents is approximately 200 times this scale. The other three datasets also exceed their break-even points at their original survey scales, by factors ranging from 2.9 to 21.9. Appendix~\ref{sec:offline_amortization} provides the exact accounting and conversion.

\phantomsection\label{sec:rq2_tradeoff}
\header{Cost-Accuracy Tradeoff.} Figure~\ref{fig:cost_accuracy_tradeoff} compares three performance curves: Qwen3-8B with different history budgets, Qwen3-32B with different history budgets, and \model\ under different values of $\alpha$. The first two curves represent fixed model--history configurations, whereas the third traces the configurations produced by \model\ as the cost preference changes. Across the evaluated values of $\alpha$, \model\ occupies the upper-right region relative to the two fixed-model curves, indicating a better Cost--Accuracy trade-off than either fixed approach. Moreover, varying $\alpha$ provides a mechanism for controlling the effectiveness--efficiency trade-off: larger values place greater emphasis on reducing inference Cost, while smaller values prioritize higher Accuracy. Thus, \model\ can operate at different points along the trade-off curve instead of being restricted to a single $\alpha$ configuration.

\subsection{Robustness Analysis}
\label{sec:rq3_robustness}

\textbf{History-Budget-Space Size.} We first test whether \model\ is robust to the set of available history budgets. Table~\ref{tab:generalization_fewshot} compares four nested action spaces, $\mathcal{K}=\{0\}$, $\{0,1\}$, $\{0,1,3\}$, and $\{0,1,3,5\}$, using Qwen3-8B and Qwen3-32B as candidate models with $\alpha=0.05$. Across both datasets and all four spaces, \model\ consistently achieves higher Accuracy and \actscore\ and lower normalized Cost than BF. In particular, the \actscore\ improvement ranges from 1.7 to 4.3 percentage points. \model\ outperforms BF even in the no-history setting $\mathcal{K}=\{0\}$, where the policy adapts only the model choice; this advantage persists as additional history budgets are introduced. This consistency indicates that the effectiveness of adaptive allocation is not tied to a particular history-budget-space definition.

\definecolor{positive}{HTML}{3F8F5F}   
\definecolor{negative}{HTML}{757575}   
\definecolor{neutral}{HTML}{757575}    
\providecommand{\posresult}[1]{\textcolor{positive}{\textbf{#1}}}
\providecommand{\negresult}[1]{\textcolor{negative}{\textbf{#1}}}
\providecommand{\neuresult}[1]{\textcolor{neutral}{\textbf{#1}}}

\begin{table*}[!t]
  \centering
  \caption{Performance across four history-budget-space sizes. Improved reports the difference between \model\ and the strongest baseline by ACT.}
  \label{tab:generalization_fewshot}
  \vspace{-0.6em}
  \setlength{\tabcolsep}{4.2pt}
  \renewcommand{\arraystretch}{1.08}
  \resizebox{\textwidth}{!}{%
    \begin{tabular}{llcccccccccccc}
      \toprule
      \multirow{2}{*}{\textbf{Dataset}}
      & \multirow{2}{*}{\textbf{Method}}
      & \multicolumn{3}{c}{$\boldsymbol{\mathcal{K}=\{0\}}$}
      & \multicolumn{3}{c}{$\boldsymbol{\mathcal{K}=\{0,1\}}$}
      & \multicolumn{3}{c}{$\boldsymbol{\mathcal{K}=\{0,1,3\}}$}
      & \multicolumn{3}{c}{$\boldsymbol{\mathcal{K}=\{0,1,3,5\}}$} \\
      \cmidrule(lr){3-5}\cmidrule(lr){6-8}\cmidrule(lr){9-11}\cmidrule(lr){12-14}
      &
      & \textbf{Accuracy} & \textbf{Cost} & \textbf{ACT}
      & \textbf{Accuracy} & \textbf{Cost} & \textbf{ACT}
      & \textbf{Accuracy} & \textbf{Cost} & \textbf{ACT}
      & \textbf{Accuracy} & \textbf{Cost} & \textbf{ACT} \\
      \midrule
      \multirow{6}{*}{ANES} & WCR & 0.3068 & 0.7106 & 0.2712 & 0.2425 & 0.6471 & 0.2101 & 0.2150 & 0.5555 & 0.1872 & 0.2047 & 0.5559 & 0.1770  \\
      & Random & 0.4355 & 0.5762 & 0.4067 & 0.5245 & 0.5016 & 0.4994 & 0.5567 & 0.4083 & 0.5363 & 0.5737 & 0.3917 & 0.5542  \\
      & LLM-S & 0.4485 & 2.3777 & 0.3296 & 0.6040 & 2.1173 & 0.4981 & 0.6150 & 1.6820 & 0.5309 & 0.6360 & 1.5882 & 0.5566  \\
      & BF & 0.4480 & 0.7485 & 0.4106 & 0.6185 & 0.7263 & 0.5822 & 0.6552 & 0.6756 & 0.6215 & 0.6552 & 0.5736 & 0.6266  \\
      & Ours & 0.4652 & 0.4886 & 0.4408 & 0.6412 & 0.4503 & 0.6187 & 0.6753 & 0.3775 & 0.6564 & 0.6880 & 0.3614 & 0.6699  \\
      & Improved & \posresult{0.0172} & \posresult{0.2599} & \posresult{0.0302} & \posresult{0.0227} & \posresult{0.2760} & \posresult{0.0365} & \posresult{0.0201} & \posresult{0.2981} & \posresult{0.0349} & \posresult{0.0328} & \posresult{0.2122} & \posresult{0.0433}  \\
      \midrule
      \multirow{6}{*}{BSA} & WCR & 0.2165 & 0.6420 & 0.1844 & 0.1375 & 0.6094 & 0.1070 & 0.1175 & 0.5897 & 0.0880 & 0.1092 & 0.5871 & 0.0799  \\
      & Random & 0.3683 & 0.5288 & 0.3418 & 0.4010 & 0.4644 & 0.3778 & 0.4145 & 0.4083 & 0.3941 & 0.4275 & 0.3848 & 0.4083  \\
      & LLM-S & 0.3593 & 2.2453 & 0.2470 & 0.4377 & 1.9700 & 0.3393 & 0.4490 & 1.6337 & 0.3673 & 0.4675 & 1.5262 & 0.3912  \\
      & BF & 0.3867 & 0.6660 & 0.3535 & 0.4597 & 0.6767 & 0.4259 & 0.4825 & 0.7046 & 0.4473 & 0.4825 & 0.5680 & 0.4541  \\
      & Ours & 0.3960 & 0.5075 & 0.3706 & 0.4855 & 0.4693 & 0.4620 & 0.5100 & 0.3929 & 0.4904 & 0.5102 & 0.3870 & 0.4909  \\
      & Improved & \posresult{0.0093} & \posresult{0.1585} & \posresult{0.0171} & \posresult{0.0258} & \posresult{0.2074} & \posresult{0.0361} & \posresult{0.0275} & \posresult{0.3117} & \posresult{0.0431} & \posresult{0.0277} & \posresult{0.1810} & \posresult{0.0368}  \\
      \bottomrule
    \end{tabular}%
  }
\end{table*}

\textbf{Model-Pool Size.} We next vary the number of candidate models while fixing $\mathcal{K}=\{0,1,3,5\}$ and $\alpha=0.05$. Table~\ref{tab:generalization_model_pool} considers a single-model pool containing Qwen3-8B and a heterogeneous three-model pool containing Qwen3-8B, Qwen3-32B, and DeepSeek-V4-Flash. With one model, \model\ can adapt only the history budget. Relative to BF, it improves Accuracy by 6.3 and 3.0 percentage points, reduces Cost by 11.6\% and 18.4\%, and raises \actscore\ by 6.6 and 3.5 percentage points on ANES and BSA, respectively. With three models, where both the model and history budget can be selected, \model\ improves Accuracy by 2.0 and 1.1 percentage points, reduces Cost by 38.9\% and 31.3\%, and raises \actscore\ by 3.1 and 1.9 percentage points on ANES and BSA, respectively. Thus, \model\ retains its effectiveness--efficiency advantage under both restricted and expanded model pools.

\definecolor{positive}{HTML}{3F8F5F}   
\definecolor{negative}{HTML}{757575}   
\definecolor{neutral}{HTML}{757575}    
\providecommand{\posresult}[1]{\textcolor{positive}{\textbf{#1}}}
\providecommand{\negresult}[1]{\textcolor{negative}{\textbf{#1}}}
\providecommand{\neuresult}[1]{\textcolor{neutral}{\textbf{#1}}}

\begin{table*}[!t]
  \centering
  \caption{Performance across two model-pool sizes. Improved reports the difference between \model\ and the strongest baseline by ACT.}
  \label{tab:generalization_model_pool}
  \vspace{-0.6em}
  \setlength{\tabcolsep}{4.2pt}
  \renewcommand{\arraystretch}{1.08}
  \resizebox{\textwidth}{!}{%
    \begin{tabular}{llcccccccccccc}
      \toprule
      \multirow{2}{*}{\textbf{Dataset}}
      & \multirow{2}{*}{\textbf{Metric}}
      & \multicolumn{6}{c}{\textbf{One Model}}
      & \multicolumn{6}{c}{\textbf{Three Models}} \\
      \cmidrule(lr){3-8}\cmidrule(lr){9-14}
      & & \textbf{WCR} & \textbf{Random} & \textbf{LLM-S} & \textbf{BF} & \textbf{Ours} & \textbf{Improved}
      & \textbf{WCR} & \textbf{Random} & \textbf{LLM-S} & \textbf{BF} & \textbf{Ours} & \textbf{Improved} \\
      \midrule
      \multirow{3}{*}{ANES} & Accuracy & 0.2752 & 0.5510 & 0.5897 & 0.5945 & 0.6575 & \posresult{0.0630} & 0.1810 & 0.5917 & 0.6522 & 0.6850 & 0.7053 & \posresult{0.0203}  \\
      & Cost & 0.6213 & 0.5206 & 2.5468 & 0.5736 & 0.5071 & \posresult{0.0665} & 0.5727 & 0.4339 & 1.8424 & 0.5622 & 0.3434 & \posresult{0.2188}  \\
      & ACT & 0.2442 & 0.5250 & 0.4624 & 0.5658 & 0.6321 & \posresult{0.0663} & 0.1524 & 0.5701 & 0.5601 & 0.6569 & 0.6881 & \posresult{0.0312}  \\
      \midrule
      \multirow{3}{*}{BSA} & Accuracy & 0.1792 & 0.4095 & 0.4507 & 0.4435 & 0.4730 & \posresult{0.0295} & 0.0902 & 0.4585 & 0.4760 & 0.5155 & 0.5262 & \posresult{0.0107}  \\
      & Cost & 0.6394 & 0.5106 & 2.4984 & 0.5680 & 0.4633 & \posresult{0.1047} & 0.6095 & 0.4223 & 1.7778 & 0.5486 & 0.3768 & \posresult{0.1718}  \\
      & ACT & 0.1473 & 0.3840 & 0.3258 & 0.4151 & 0.4498 & \posresult{0.0347} & 0.0598 & 0.4374 & 0.3871 & 0.4881 & 0.5074 & \posresult{0.0193}  \\
      \bottomrule
    \end{tabular}%
  }
\end{table*}

\begin{figure*}[!t]
    \centering
    \includegraphics[width=\textwidth]{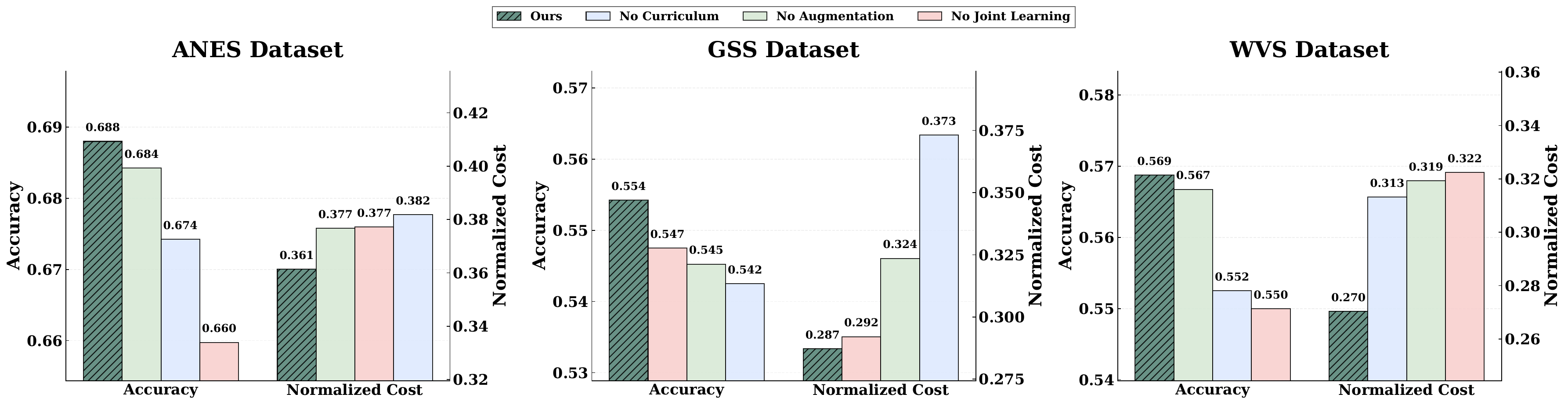}
    \caption{Ablation study of \model\ on ANES, GSS, and WVS. We compare the full framework with variants removing the curriculum, respondent-history augmentation, or joint model--history selection. The left and right axes show Accuracy and Normalized Cost, respectively.}
    \label{fig:ablation_study}
    \vspace{-1em}
\end{figure*}

\subsection{Ablation Study}
\label{sec:rq4_ablation}

We conduct ablation studies to examine the contribution of the \actscore-margin curriculum, respondent-history augmentation, and joint model--history selection in \model. All experiments use Qwen3-8B and Qwen3-32B, the history-budget space $\mathcal{K}=\{0,1,3,5\}$, and $\alpha=0.05$. Figure~\ref{fig:ablation_study} reports the results on ANES, GSS, and WVS. We evaluate the following variants:

\begin{itemize}[leftmargin=*, itemsep=2pt, topsep=2pt]
    \item[\ablationdot{AblationBlue}] No Curriculum: removes the \actscore-margin curriculum and trains on all requests immediately.
    \item[\ablationdot{AblationGreen}] No Augmentation: disables historical-response deletion and permutation during training.
    \item[\ablationdot{AblationPink}] No Joint Learning: replaces joint model--history selection with a two-stage procedure that first selects the model and then selects the history budget (Appendix~\ref{sec:two_stage_ablation}).
\end{itemize}

As shown in Figure~\ref{fig:ablation_study}, the full \model\ consistently
achieves higher Accuracy and lower Cost than all three ablated variants across
the three datasets. Removing the curriculum causes the largest degradation on
GSS, while replacing joint selection with the two-stage procedure has the
greatest impact on ANES and WVS. Removing respondent-history augmentation also leads to
lower Accuracy and higher Cost on every dataset. These results demonstrate
that all three components contribute to the effectiveness--efficiency
trade-off of \model.





\section{Related Work}
\label{sec:related_work}
We review two lines of work most relevant to our study: LLM-based survey simulation and LLM Routing and Adaptive Inference.


\textbf{LLM-based Survey Simulation.}
Large language models have increasingly been used as synthetic respondents in survey research. Early studies showed that LLMs conditioned on sociodemographic profiles or persona descriptions can reproduce aspects of human attitudes and response distributions, motivating their use for scalable survey simulation~\citep{argyle2023out,aher2023using,durmus2023towards,cao2023assessing}. Subsequent work further examined whose opinions LLMs reflect and the limitations of using synthetic responses as substitutes for human survey data~\citep{santurkar2023whose,bisbee2024synthetic}. More recently, research has moved toward specialized models, richer respondent conditioning, and more systematic evaluation~\citep{chen2026benchmarking}. These include fine-tuning models to reproduce population-level response distributions~\citep{cao2025specializing,suh2025language,huang2026distribution}, as well as respondent-level simulation using sociodemographic personas, personality information, and historical survey responses~\citep{wang2025sociobench,lutz2025prompt,liu2025beyond,ku2026silicon}.
Recent studies also show that survey simulation is sensitive to persona construction, attribute selection, response-generation strategies, and prompt design, and that providing more respondent information does not necessarily improve fidelity~\citep{lutz2025prompt,morocho2026assessing,ahnert2026survey,frohling2026persona}. However, these methods generally treat the model and respondent context as study-level configuration choices. In contrast, \model\ treats them as request-dependent resources and jointly allocates model capability and historical-response context for each respondent--question pair under an explicit accuracy--cost objective.


\textbf{LLM Routing and Adaptive Inference.}
LLM routing improves inference efficiency by dynamically assigning each query to a model with an appropriate capability--cost trade-off. Existing approaches use signals such as preference feedback, query representations, contrastive supervision, or model--query relations to learn routing policies, while cascading methods progressively invoke stronger models when additional capability is needed~\citep{ong2025routellmlearningroutellms,mohammadshahi2024routoo,chen2024routerdc,feng2024graphrouter,dekoninck2024unified}. Recent work further studies budget-aware online routing and large-scale evaluation of routing policies~\citep{panda2025adaptive,huang2025routereval}.
More broadly, our work considers adaptive inference as a resource-allocation problem rather than model selection alone. In survey simulation, different respondent--question requests may require different levels of model capability and different amounts of respondent history. \model\ therefore jointly allocates these two resources at the request level, selecting both the model and the history budget under an explicit accuracy--cost objective.

\section{Conclusion}
\label{sec:conclusion}

In this paper, we identified the limitations of fixed configurations in LLM-based social survey simulation by examining the request-dependent effects of model capability, respondent history, and their interaction. To address this problem, we proposed \model, an adaptive resource allocation framework that jointly selects the model and history budget for each request. The framework combines \actscore-margin curriculum learning with respondent-history drop-and-swap augmentation. Across the evaluated settings, \model\ generally achieves higher observed \actscore\ than the compared baselines while improving Accuracy and reducing inference Cost in most cases.
Future work will extend the action space and adapt model-cost weights to deployment-specific pricing (Appendix~\ref{sec:limitations}).

\section*{AI use statement}
Language models were used as experimental components in this work: the
candidate LLMs generated simulated survey responses, and Qwen3-32B served as
the LLM-based configuration-selection baseline. Separately, generative AI
tools provided limited assistance with manuscript
language editing, translation, formatting, and software troubleshooting. They
were not used to originate the research idea, formulate the method, design the
experiments, alter the datasets or reported results, or interpret the
findings. All AI-assisted edits and suggestions were reviewed by the authors,
who take responsibility for the final manuscript and all of its claims, code,
and other artifacts.

\section*{Ethics statement}
This work uses publicly released social-survey datasets and publicly available
language models, subject to their respective terms of use and licenses. We do
not collect new data from human participants. The simulated responses are
research artifacts and should not be treated as substitutes for informed
responses from real individuals or as authoritative statements about any
demographic group. Because language-model outputs can reproduce biases or
errors present in their training data and prompts, the reported simulations
should be interpreted with appropriate caution.

\section*{Reproducibility statement}
We provide the formal problem definition and training procedure in
Sections~\ref{sec:problem_formulation} and~\ref{sec:method}, the experimental
settings and evaluation protocol in Section~\ref{sec:experiments}, and the
evaluation prompt templates in Appendix~\ref{sec:prompt_templates}. The source
code, experiment launchers, figures, and table-generation scripts are released
at \url{https://anonymous.4open.science/r/E2Sim-DE66}. Exact reproduction additionally
requires access to the survey data specified in the code, model endpoints or
checkpoints, the embedding model, and the documented model-price settings.

\bibliographystyle{assets/plainnat}
\bibliography{paper}

@article{achiam2023gpt,
  title={Gpt-4 technical report},
  author={Achiam, Josh and Adler, Steven and Agarwal, Sandhini and Ahmad, Lama and Akkaya, Ilge and Aleman, Florencia Leoni and Almeida, Diogo and Altenschmidt, Janko and Altman, Sam and Anadkat, Shyamal and others},
  journal={arXiv preprint arXiv:2303.08774},
  year={2023}
}

@article{kalton2009methods,
  title={Methods for oversampling rare subpopulations in social surveys},
  author={Kalton, Graham},
  journal={Survey methodology},
  volume={35},number={2},
  pages={125--141},
  year={2009}
}

@article{heffetz2019difficulty,
  title={Difficulty of reaching respondents and nonresponse Bias: Evidence from large government surveys},
  author={Heffetz, Ori and Reeves, Daniel B},
  journal={Review of Economics and Statistics},
  volume={101},
  number={1},
  pages={176--191},
  year={2019},
  publisher={MIT Press One Rogers Street, Cambridge, MA 02142-1209, USA journals-info~…}
}

@article{wright2010survey,
  title={Survey research and social science: History, current practice, and future prospects},
  author={Wright, James D and Marsden, Peter V and others},
  journal={Handbook of survey research},
  pages={3--26},
  year={2010}
}

@book{groves2011survey,
  title={Survey methodology},
  author={Groves, Robert M and Fowler Jr, Floyd J and Couper, Mick P and Lepkowski, James M and Singer, Eleanor and Tourangeau, Roger},
  year={2011},
  publisher={John Wiley \& Sons}
}

@article{roopa2012questionnaire,
  title={Questionnaire designing for a survey},
  author={Roopa, Siddegowda and Rani, Menta Satya},
  journal={Journal of Indian Orthodontic Society},
  volume={46},
  number={4\_suppl1},
  pages={273--277},
  year={2012},
  publisher={SAGE Publications Sage India: New Delhi, India}
}

@misc{tourangeau2000psychology,
  title={The Psychology of Survey Response. Cambridge University Press; Cambridge Core},
  author={Tourangeau, R and Rips, LJ and Rasinski, K},
  year={2000}
}

@article{argyle2023out,
  title={Out of one, many: Using language models to simulate human samples},
  author={Argyle, Lisa P and Busby, Ethan C and Fulda, Nancy and Gubler, Joshua R and Rytting, Christopher and Wingate, David},
  journal={Political Analysis},
  volume={31},
  number={3},
  pages={337--351},
  year={2023},
  publisher={Cambridge University Press}
}

@inproceedings{aher2023using,
  title={Using large language models to simulate multiple humans and replicate human subject studies},
  author={Aher, Gati V and Arriaga, Rosa I and Kalai, Adam Tauman},
  booktitle={International conference on machine learning},
  pages={337--371},
  year={2023},
  organization={PMLR}
}

@inproceedings{santurkar2023whose,
  title={Whose opinions do language models reflect?},
  author={Santurkar, Shibani and Durmus, Esin and Ladhak, Faisal and Lee, Cinoo and Liang, Percy and Hashimoto, Tatsunori},
  booktitle={International conference on machine learning},
  pages={29971--30004},
  year={2023},
  organization={PMLR}
}

@article{bisbee2024synthetic,
  title={Synthetic replacements for human survey data? The perils of large language models},
  author={Bisbee, James and Clinton, Joshua D and Dorff, Cassy and Kenkel, Brenton and Larson, Jennifer M},
  journal={Political Analysis},
  volume={32},
  number={4},
  pages={401--416},
  year={2024},
  publisher={Cambridge University Press}
}

@inproceedings{cao2023assessing,
  title={Assessing cross-cultural alignment between ChatGPT and human societies: An empirical study},
  author={Cao, Yong and Zhou, Li and Lee, Seolhwa and Piqueras, Laura Cabello and Chen, Min and Hershcovich, Daniel},
  booktitle={Proceedings of the first workshop on cross-cultural considerations in NLP (C3NLP)},
  pages={53--67},
  year={2023}
}

@article{durmus2023towards,
  title={Towards measuring the representation of subjective global opinions in language models},
  author={Durmus, Esin and Nguyen, Karina and Liao, Thomas I and Schiefer, Nicholas and Askell, Amanda and Bakhtin, Anton and Chen, Carol and Hatfield-Dodds, Zac and Hernandez, Danny and Joseph, Nicholas and others},
  journal={arXiv preprint arXiv:2306.16388},
  year={2023}
}

@article{squazzoni2020computational,
  title={Computational models that matter during a global pandemic outbreak: A call to action},
  author={Squazzoni, Flaminio and Polhill, J Gareth and Edmonds, Bruce and Ahrweiler, Petra and Antosz, Patrycja and Scholz, Geeske and Chappin, {\'E}mile and Borit, Melania and Verhagen, Harko and Giardini, Francesca and others},
  journal={Jasss},
  volume={23},
  number={2},
  year={2020},
  publisher={University of Surrey Department of Sociology (JASS S.)}
}

@article{hamill2009social,
  title={Social circles: A simple structure for agent-based social network models},
  author={Hamill, Lynne and Gilbert, Geoffrey},
  journal={Journal of Artificial Societies and Social Simulation},
  volume={12},
  number={2},
  year={2009},
  publisher={SimSoc Consortium}
}

@article{wang2025user,
  title={User behavior simulation with large language model-based agents},
  author={Wang, Lei and Zhang, Jingsen and Yang, Hao and Chen, Zhi-Yuan and Tang, Jiakai and Zhang, Zeyu and Chen, Xu and Lin, Yankai and Sun, Hao and Song, Ruihua and others},
  journal={ACM Transactions on Information Systems},
  volume={43},
  number={2},
  pages={1--37},
  year={2025},
  publisher={ACM New York, NY}
}

@article{chen2026benchmarking,
  title={Benchmarking LLMs for Community Governance Simulation with Life-history Narratives},
  author={Chen, Xu and Li, Yuanzi and Wang, Lei and Lu, Nan and Wang, Yang and Wang, Anding and Shi, Lei and Fu, Xiaoxing and Wen, Ji-Rong},
  journal={arXiv preprint arXiv:2605.23783},
  year={2026}
}

@article{mohammadshahi2024routoo,
  title={Routoo: Learning to route to large language models effectively},
  author={Mohammadshahi, Alireza and Shaikh, Arshad Rafiq and Yazdani, Majid},
  journal={arXiv preprint arXiv:2401.13979},
  year={2024}
}

@article{chen2024routerdc,
  title={Routerdc: Query-based router by dual contrastive learning for assembling large language models},
  author={Chen, Shuhao and Jiang, Weisen and Lin, Baijiong and Kwok, James and Zhang, Yu},
  journal={Advances in Neural Information Processing Systems},
  volume={37},
  pages={66305--66328},
  year={2024}
}

@article{feng2024graphrouter,
  title={Graphrouter: A graph-based router for llm selections},
  author={Feng, Tao and Shen, Yanzhen and You, Jiaxuan},
  journal={arXiv preprint arXiv:2410.03834},
  year={2024}
}

@article{huang2025routereval,
  title={Routereval: A comprehensive benchmark for routing llms to explore model-level scaling up in llms},
  author={Huang, Zhongzhan and Ling, Guoming and Lin, Yupei and Chen, Yandong and Zhong, Shanshan and Wu, Hefeng and Lin, Liang},
  journal={arXiv preprint arXiv:2503.10657},
  year={2025}
}

@article{frohling2026persona,
  title={When Persona Attributes Improve Population Alignment in Large Language Models},
  author={Fr{\"o}hling, Leon and Rupprecht, Jens and Strohmaier, Markus and Wagner, Claudia},
  journal={arXiv preprint arXiv:2609.02526},
  year={2026}
}

@inproceedings{ahnert2026survey,
  title={Survey response generation: Generating closed-ended survey responses in-silico with large language models},
  author={Ahnert, Georg and Haensch, Anna-Carolina and Plank, Barbara and Strohmaier, Markus},
  booktitle={Proceedings of the 64th Annual Meeting of the Association for Computational Linguistics (Volume 1: Long Papers)},
  pages={41554--41577},
  year={2026}
}

@article{morocho2026assessing,
  title={Assessing the Reliability of Persona-Conditioned LLMs as Synthetic Survey Respondents},
  author={Morocho, Erika Elizabeth Taday and Cima, Lorenzo and Fagni, Tiziano and Avvenuti, Marco and Cresci, Stefano},
  journal={arXiv preprint arXiv:2602.18462},
  year={2026}
}

@article{ku2026silicon,
  title={Silicon Sampling via Cross-Survey Transfer},
  author={Ku, Chan-Tung and Hsu, Chan and Huang, Pei-Cing and Liu, Frank Cheng-shan and Cheng, I and Kang, Yihuang and others},
  journal={arXiv preprint arXiv:2607.03091},
  year={2026}
}

@inproceedings{huang2026distribution,
  title={Distribution shift alignment helps LLMs simulate survey response distributions},
  author={Huang, Ji and Mengfei, LI and Shao, Shuai},
  booktitle={Findings of the Association for Computational Linguistics: ACL 2026},
  pages={9395--9409},
  year={2026}
}

@inproceedings{liu2025beyond,
  title={Beyond demographics: Enhancing cultural value survey simulation with multi-stage personality-driven cognitive reasoning},
  author={Liu, Haijiang and Li, Qiyuan and Gao, Chao and Cao, Yong and Xu, Xiangyu and Wu, Xun and Hershcovich, Daniel and Gu, Jinguang},
  booktitle={Proceedings of the 2025 Conference on Empirical Methods in Natural Language Processing},
  pages={18417--18439},
  year={2025}
}

@inproceedings{lutz2025prompt,
  title={The Prompt Makes the Person (a): A Systematic Evaluation of Sociodemographic Persona Prompting for Large Language Models.},
  author={Lutz, Marlene and Sen, Indira and Ahnert, Georg and Rogers, Elisa and Strohmaier, Markus},
  booktitle={EMNLP (Findings)},
  pages={23212--23237},
  year={2025}
}

@inproceedings{wang2025sociobench,
  title={Sociobench: Modeling human behavior in sociological surveys with large language models},
  author={Wang, Jia and Zhao, Ziyu and Ni, Tingjuntao and Wei, Zhongyu},
  booktitle={Proceedings of the 2025 Conference on Empirical Methods in Natural Language Processing},
  pages={26268--26300},
  year={2025}
}

@inproceedings{suh2025language,
  title={Language model fine-tuning on scaled survey data for predicting distributions of public opinions},
  author={Suh, Joseph and Jahanparast, Erfan and Moon, Suhong and Kang, Minwoo and Chang, Serina},
  booktitle={Proceedings of the 63rd Annual Meeting of the Association for Computational Linguistics (Volume 1: Long Papers)},
  pages={21147--21170},
  year={2025}
}

@inproceedings{cao2025specializing,
  title={Specializing large language models to simulate survey response distributions for global populations},
  author={Cao, Yong and Liu, Haijiang and Arora, Arnav and Augenstein, Isabelle and R{\"o}ttger, Paul and Hershcovich, Daniel},
  booktitle={Proceedings of the 2025 Conference of the Nations of the Americas Chapter of the Association for Computational Linguistics: Human Language Technologies (Volume 1: Long Papers)},
  pages={3141--3154},
  year={2025}
}

@inproceedings{panda2025adaptive,
  title={Adaptive LLM Routing under Budget Constraints.},
  author={Panda, Pranoy and Magazine, Raghav and Devaguptapu, Chaitanya and Takemori, Sho and Sharma, Vishal},
  booktitle={EMNLP (Findings)},
  pages={23934--23949},
  year={2025}
}

@article{dekoninck2024unified,
  title={A unified approach to routing and cascading for llms},
  author={Dekoninck, Jasper and Baader, Maximilian and Vechev, Martin},
  journal={arXiv preprint arXiv:2410.10347},
  year={2024}
}

@article{minaee2024large,
  title={Large language models: A survey},
  author={Minaee, Shervin and Mikolov, Tomas and Nikzad, Narjes and Chenaghlu, Meysam and Socher, Richard and Amatriain, Xavier and Gao, Jianfeng},
  journal={arXiv preprint arXiv:2402.06196},
  year={2024}
}

@misc{ong2025routellmlearningroutellms,
      title={RouteLLM: Learning to Route LLMs with Preference Data}, 
      author={Isaac Ong and Amjad Almahairi and Vincent Wu and Wei-Lin Chiang and Tianhao Wu and Joseph E. Gonzalez and M Waleed Kadous and Ion Stoica},
      year={2025},
      eprint={2406.18665},
      archivePrefix={arXiv},
      primaryClass={cs.LG},
      url={https://arxiv.org/abs/2406.18665}, 
}
\clearpage
\appendix


\section{Dataset and Evaluation Protocols}
\label{app:data_splits}

\subsection{Dataset Statistics}

For each survey, we organize the observed responses as a respondent--question matrix and use 20,000 observed entries as the experimental pool. Each entry is a respondent--question pair together with its recorded response. Table~\ref{tab:dataset_split_statistics} reports the numbers of respondents and questions in each pool and the training and test sizes. We randomly assign 20\% of observed pairs to test using split seed 0. For both training and test requests, history is retrieved only from
training responses, excluding the target pair; test answers are used solely
for evaluation.

\begin{table}[H]
    \centering
    \footnotesize
    \setlength{\tabcolsep}{7pt}
    \caption{Dataset statistics and training/test split sizes.}
    \label{tab:dataset_split_statistics}
    \vspace{-0.6em}
    \begin{tabular}{lrrrrr}
        \toprule
        Dataset & Pairs & Respondents & Questions & Train & Test \\
        \midrule
        ANES & 20,000 & 88 & 332 & 16,000 & 4,000 \\
        BSA & 20,000 & 430 & 119 & 16,000 & 4,000 \\
        GSS & 20,000 & 265 & 208 & 16,000 & 4,000 \\
        WVS & 20,000 & 157 & 175 & 16,000 & 4,000 \\
        \bottomrule
    \end{tabular}
\end{table}

\section{Prompt Templates}
\label{sec:prompt_templates}

This section records the prompt templates used for survey-response generation
and LLM-based configuration selection.

\subsection{Survey-Response Generation Prompt}

The following template is used to generate $\hat{y}_i(a)$
for every model--history configuration (Eq.~\ref{eq:action_response}). The
placeholders in angle brackets are filled with the respondent persona,
retrieved historical responses, and target question for each request.

\begin{PromptBox}{Complete system and user prompt template}
SYSTEM PROMPT:
You are simulating the survey respondent described by the persona. Answer the target survey question as that respondent would answer it. Use past answers only as behavioral evidence. Select exactly one listed option and return only a JSON object in the form {"answer": <integer>}.

USER PROMPT:
PERSONA:
<respondent persona>

PAST ANSWERS (most similar first):
Example 1 (semantic similarity=<score>):
<historical question text>
Options:
[0] <option 0>
[1] <option 1>
...
Respondent's answer: [<answer index>] <answer text>

Example 2 (semantic similarity=<score>):
<historical question text>
Options:
[0] <option 0>
[1] <option 1>
...
Respondent's answer: [<answer index>] <answer text>

<additional examples, if any>

TARGET QUESTION:
<target question text>
Options:
[0] <option 0>
[1] <option 1>
...
Return only: {"answer": <integer>}
\end{PromptBox}

\subsection{LLM-based Selector Prompt}
\label{sec:llm_selector_prompt}

This section records the exact prompt used by the LLM-based Selector (LLM-S)
baseline described in Section~\ref{sec:experiments}. Given the respondent
persona, target question, and the candidate
model--history configurations with their normalized costs, the selector
returns the index of one action $a\in\mathcal{A}$.

\begin{PromptBox}{Complete LLM-based Selector system and user prompt}
SYSTEM PROMPT:
You are an expert system that selects the optimal language model configuration for survey response prediction.

Given:
1. A respondent persona
2. A survey question to answer
3. Available model configurations with their normalized costs

Your task is to maximize the REWARD function:
  Reward = Accuracy - 0.05 x Cost

Where:
- Accuracy: prediction correctness (0 or 1)
- Cost: normalized computational cost of the selected configuration
- Alpha (0.05): cost penalty weight

This means:
- A one-percentage-point accuracy improvement offsets an increase of 0.2 in cost (0.01 / 0.05 = 0.2)
- Accuracy is the dominant factor, but cost matters
- Avoid unnecessarily expensive configurations when simpler ones suffice
- Don't sacrifice accuracy just to save cost

Return ONLY a JSON object in the format: {"selected_action": <action_index>}

USER PROMPT:
RESPONDENT PERSONA:
<respondent persona>

TARGET SURVEY QUESTION:
<target question with its options>

AVAILABLE RESPONDENT HISTORY (retrieved similar questions):
  Example 1:
  <example question with its options, if available>
  Respondent's answer: [<answer index>] <answer text>

  Example 2:
  <example question with its options, if available>
  Respondent's answer: [<answer index>] <answer text>

  <additional questions and respondent answers, if available>

AVAILABLE CONFIGURATIONS:
Action 0: Model=<model>, history budget=0 (no history), Normalized Cost=<normalized_cost>
Action 1: Model=<model>, history budget=<k> (use first <k> responses above), Normalized Cost=<normalized_cost>
<one line for every candidate action>

OPTIMIZATION OBJECTIVE:
Maximize: Reward = Accuracy - 0.05 x Cost

DECISION STRATEGY:
1. Estimate the expected accuracy for each configuration based on:
   - Question complexity (factual vs. opinion-based, simple vs. nuanced)
   - Persona characteristics (education, political engagement, demographic factors)
   - Model capability (larger models handle complex reasoning better)
   - history relevance (do the responses help with this specific question?)

2. Evaluate the cost-benefit tradeoff:
   - Use the normalized Cost values listed above
   - Prefer a higher-Cost configuration only when its expected Accuracy gain exceeds 0.05 times the Cost increase

3. Apply heuristics:
   - Simple factual questions -> cheaper configs often sufficient
   - Complex political/value questions -> larger models often worth the cost
   - Highly educated/engaged personas -> may need more nuanced predictions
   - Generic questions with clear answers -> respondent history may not help much

Select the action index that maximizes the expected reward.
\end{PromptBox}


\section{Cost Computation}
\label{sec:cost_computation}

For a selected survey action $a=(m,k)$ on request $i$, let $T_{i,a}$ denote
the API-reported prompt-token count and let $c_m^{\mathrm{in}}$ denote the
relative input-token price of model $m$. The execution cost is
\begin{equation}
 C^{\mathrm{exec}}_i=c_m^{\mathrm{in}}T_{i,a}.
\end{equation}
Both LLM-based Selector and \model\ incur an additional configuration-selection
cost before this execution step, since each request must first be routed to an
action $a$. For LLM-based Selector, let $T_i^{\mathrm{selector}}$ be the selector
prompt-token count and $c_{\mathrm{selector}}^{\mathrm{in}}$ its price multiplier
(the Qwen3-32B input-token price). The selector cost is
\begin{equation}
 C^{\mathrm{selector}}_i=c_{\mathrm{selector}}^{\mathrm{in}}T_i^{\mathrm{selector}}.
\end{equation}
For \model, let $T_i^{\mathrm{policy}}$ be the input-token count of the policy
input $z(x)$ defined in Section~\ref{sec:inference} and fed to the ModernBERT-base policy
model, priced at $c_{\mathrm{policy}}^{\mathrm{in}}=2\times149\text{M}/8\text{B}\approx0.0373$
on the same relative scale as Qwen3-8B, by parameter count.
The policy-model cost is
\begin{equation}
 C^{\mathrm{policy}}_i=c_{\mathrm{policy}}^{\mathrm{in}}T_i^{\mathrm{policy}}.
\end{equation}
Let $C_i^{\mathrm{select}}$ denote the applicable selection cost for a given
method: $C_i^{\mathrm{select}}=C_i^{\mathrm{selector}}$ for LLM-based Selector,
$C_i^{\mathrm{select}}=C_i^{\mathrm{policy}}$ for \model, and
$C_i^{\mathrm{select}}=0$ for methods without a selection step (WCR, Random,
Oracle Best Fixed). The average total cost of a method is
\begin{equation}
 \overline{C}^{\mathrm{total}}=\frac{1}{N}\sum_{i=1}^{N}
 \left(C^{\mathrm{exec}}_i+C^{\mathrm{select}}_i\right).
\end{equation}
To ensure comparability,
the normalizer is defined once for each dataset, model pool, history-budget action
space, and evaluated request set, and the same value is used for every method
compared under that setting. It is the maximum candidate execution cost over
all evaluated requests and candidate actions:
\begin{equation}
 C_{\max}=\max_{i,\,a=(m,k)\in\mathcal A}
 c_m^{\mathrm{in}}T_{i,a},
 \qquad
 \operatorname{Cost}=\frac{\overline{C}^{\mathrm{total}}}{C_{\max}}.
\end{equation}
The reported LLM-based Selector results include the Qwen3-32B selection cost,
and the reported \model\ results include the ModernBERT-base policy-model
cost, in the numerator, so the normalized Cost can exceed one for either
method. The reported trade-off score is $\operatorname{Acc}-\alpha\operatorname{Cost}$.

\section{Offline-Supervision Amortization Details}
\label{sec:offline_amortization}

\begin{table}[t]
    \centering
    \small
    \setlength{\tabcolsep}{5.5pt}
    \caption{ACT-based amortization of offline supervision. BE denotes break-even, Scale is BE requests relative to training, and Full/BE is the original survey population relative to BE respondents.}
    \label{tab:offline_amortization}
    \vspace{-0.6em}
    \resizebox{\textwidth}{!}{%
    \begin{tabular}{lrrrrrrr}
        \toprule
        Dataset & Train Resp. & Full Resp. & Req./Resp. & BE Req. & BE Resp. & Scale & Full/BE \\
        \midrule
        ANES & 88 & 5,521 & 227.3 & 57K & 252 & 3.6$\times$ & 21.9$\times$ \\
        BSA & 430 & 4,120 & 46.5 & 66K & 1,417 & 4.1$\times$ & 2.9$\times$ \\
        GSS & 265 & 3,986 & 75.5 & 78K & 1,036 & 4.9$\times$ & 3.8$\times$ \\
        WVS & 157 & 97,220 & 127.4 & 62K & 485 & 3.9$\times$ & 200.4$\times$ \\
        \bottomrule
    \end{tabular}%
    }
    \vspace{-0.6em}
\end{table}

We measure how many deployment requests are needed to recover the one-time
cost of constructing \model's offline supervision. For the primary model pool,
each of the $N_{\mathrm{train}}=16{,}000$ training requests is evaluated under
eight actions. Using the recorded prompt-token counts and the same model-price
weights as the main experiments, the exact acquisition cost is
\begin{equation}
 C_{\mathrm{offline}}=
 \sum_{i\in\mathcal D_{\mathrm{train}}}
 \sum_{a=(m,k)\in\mathcal A}c_m^{\mathrm{in}}T(x_i,a).
\end{equation}
Let $C_{\max}$ be the shared normalizer from Section~\ref{sec:cost_computation}.
Under the paper's objective, the offline acquisition contributes the one-time
penalty $\alpha C_{\mathrm{offline}}/C_{\max}$. The per-request deployment gain
relative to BF is
\begin{equation}
 \Delta\overline g=
 \left(\overline{\operatorname{Acc}}_{\mathrm{E2Sim}}
 -\alpha\overline C_{\mathrm{E2Sim}}\right)
 -\left(\overline{\operatorname{Acc}}_{\mathrm{BF}}
 -\alpha\overline C_{\mathrm{BF}}\right).
\end{equation}
The ACT-based break-even request volume is therefore
\begin{equation}
 N_{\mathrm{BE}}=
 \frac{\alpha C_{\mathrm{offline}}/C_{\max}}
 {\Delta\overline g}.
\end{equation}
This formulation credits both the Accuracy improvement and the Cost reduction
under the same objective used to train and evaluate \model. The comparison
remains conservative toward the fixed baseline because BF is assigned no
offline selection cost. The Two-Stage Router is not used as the amortization
reference because it consumes the same exhaustive action matrix as \model, so
their acquisition costs cancel in a direct comparison. The released analysis
file additionally reports a cost-only break-even value as a stricter
sensitivity measure.

To express request volume as respondent-equivalents, we use the observed
density of the experimental pool, $\bar Q_d=20{,}000/R_d$, where $R_d$ is the
number of respondents reported in Table~\ref{tab:dataset_split_statistics}.
Thus, $R_{\mathrm{BE}}=N_{\mathrm{BE}}/\bar Q_d$. The original source surveys
contain 5,521, 4,120, 3,986, and 97,220 respondents for ANES, BSA, GSS, and
WVS, respectively. ``Full/BE'' in Table~\ref{tab:offline_amortization} divides
these population sizes by $R_{\mathrm{BE}}$; it is a workload projection that
holds the observed requests-per-respondent density fixed rather than assuming
that every respondent is simulated on every questionnaire item. Exact
intermediate values are produced from the released token logs by the
amortization-analysis script.

\section{Two-Stage Ablation Details}
\label{sec:two_stage_ablation}

\begin{figure}[H]
    \centering
    \includegraphics[width=\textwidth]{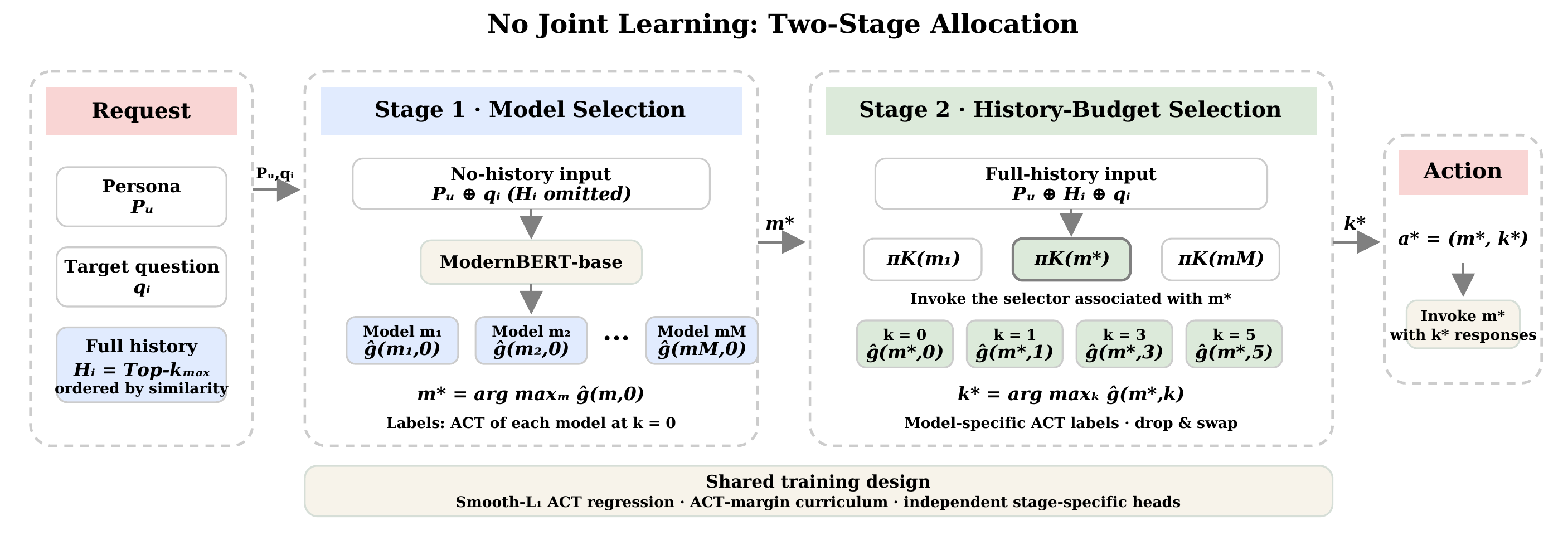}
    \caption{Two-stage allocation in the No Joint Learning ablation. Stage 1 selects a model using the predicted \actscore\ of each model's no-history action. The selector associated with the chosen model then predicts the \actscore\ of each available history budget and selects $k^*$, yielding the final action $(m^*,k^*)$.}
    \label{fig:two_stage_ablation}
\end{figure}

The No Joint Learning variant, illustrated in Figure~\ref{fig:two_stage_ablation}, uses the same ModernBERT-base architecture and Smooth-$L_1$ \actscore\ regression objective as the joint policy, but decomposes allocation into two stages. Stage 1 receives the persona and target question without respondent history and regresses the \actscore\ values of the $k=0$ action for each candidate model; the model with the highest predicted value is selected. Stage 2 trains one independent history-budget selector per model, where each selector receives the full top-$k_{\max}$ history and regresses the \actscore\ values of that model's actions over $k\in\mathcal{K}$. Both stages use the same \actscore-margin curriculum, while respondent-history drop-and-swap augmentation is applied only to the Stage 2 selectors. At inference, Stage 1 selects $m^*$ and the corresponding model-specific Stage 2 selector chooses $k^*$, producing the final action $(m^*,k^*)$.

\section{Limitations}
\label{sec:limitations}

\begin{itemize}[leftmargin=*, itemsep=2pt, topsep=2pt]

    \item[\dinglabelone] \textbf{Predefined action space.}
    \model\ operates over a fixed set of candidate models and history budgets. Introducing a new model or history budget requires evaluating the added configurations and updating the allocation policy.

    \item[\dinglabeltwo] \textbf{Empirical model costs.}
    Our experiments use empirical relative model-cost weights to provide controlled comparisons. These weights are not universal estimates of wall-clock latency or monetary cost; in deployment, they should be configured using current API prices or costs measured in the target hardware environment.

\end{itemize}

\end{document}